\documentclass[prb,superscriptaddress,floatfix,showpacs,showkeys]{revtex4}
\usepackage{amsmath,epsfig,dcolumn,natbib}

\begin{document}


%
%

\title{Numerical local \lq\lq hybrid" functional treatment of selected diatomic molecules:
comparison of energies and multipole moments to conventional hybrid functionals
}
\author{Ivan A.~Mikhailov}

\address{Nanoscience Technology Center, University of Central Florida\\
Orlando, Florida 32826, USA\\
ivan.mikhaylov@gmail.com}

\author{Olga A. Shukruto}

\address{Quantum Theory Project,
Departments of Physics and of Chemistry, P.O. Box 118435,
University of Florida, 
Gainesville, Florida 32611-8435, USA}

\author{Art\"em E. Masunov}

\address{Nanoscience Technology Center, Department of Chemistry, and Department of Physics,
University of Central Florida, 
Orlando, Florida 32826, USA}

\author{Valentin V.~Karasiev\footnote{Corresponding author.}}

\address{Quantum Theory Project,
Departments of Physics and of Chemistry, P.O. Box 118435,
University of Florida, Gainesville, Florida 32611-8435, USA\\
vkarasev@qtp.ufl.edu}



\begin{abstract}
  New local \lq\lq hybrid" functionals proposed by V. V. Karasiev in [J. Chem. Phys. {\bf 118}, 8567 (2003)] are benchmarked against nonlocal hybrid functionals. Their performance is tested on the total and high occupied orbital energies, as well as  the electric moments of selected diatomic molecules.  The new functionals, along with the Hartree-Fock and non-hybrid functionals, are employed for finite-difference  calculations, which are basis-independent.
  Basis set errors in the total energy and electric moments are calculated for the 6-311G, 6-311G++G(3df,3pd) and AUG-cc-pVnZ (n=3,4,6) 
  basis sets used in conjunction with the Hartree-Fock and conventional density functional methods.
  A comparison between the results of the finite-difference local \lq\lq hybrid" and  basis set nonlocal hybrid functional shows that total energies of local and nonlocal hybrid functionals agree to within the basis set error. Discrepancies for multipole moments are larger in magnitude when compared to the basis set errors, but still reasonably small (smaller than errors produced by the 6-311G basis set). Thus, we recommend using the new local \lq\lq hybrid" functionals whenever the accuracy is expected to be sufficient, because they require a solution of just differential Kohn-Sham equations, instead of integro-differential ones in the case of hybrid functionals.
\end{abstract}


\maketitle

\section{Introduction}

The investigation of electronic structure by means of theory represents an important
and highly active area of research. Calculations of electronic structures are
frequently performed by means of the density functional theory (DFT).\cite{KohnSham65,HohenbegrKohn,ParrYang,DreizlerGross}
DFT in the form of the Kohn-Sham (KS) method provides an approach to treat the
Schr\"odinger equation within an exact formalism that bypasses
numerically prohibitive calculations based on many-electron wavefunctions.
Within DFT, exchange and correlation effects are accounted for
by means of energy functionals of the electron density -- the exchange-correlation functionals,
and their functional derivatives with respect to the electron density -- the
exchange-correlation potentials. These functionals are approximated in
standard DFT methods.

Hybrid functionals which combine the nonlocal Hartree-Fock exchange 
with local KS exchange-correlation potentials
provide a significant improvement over conventional
DFT approaches with respect to binding energies and other properties.
Unfortunately, such hybrid methods no longer fit the framework of
the DFT formalism due to presence of a non-multiplicative operator in
corersponding one-electron equations. Moreover, the incorporation of a nonlocal Hartee-Fock
exchange operator, defined by its action on a particular electron orbital as
\begin{eqnarray}
\hat v_{\rm x}^{\rm HF}({\bf r})\psi_i({\bf r}) &=&
\frac{\delta E_{\rm x}^{\rm HF}[\{\psi_i\}]}{\delta \psi^*_i({\bf r})}
\nonumber\\
&=&
- \sum_{j=1}^{N} \delta(\sigma_i, \sigma_j)\psi_j({\bf r})
\int \frac{\psi_j^*({\bf r}')\psi_i({\bf r}')}{|{\bf r}-{\bf r}'|}d{\bf r}' ,
\label{I1}
\end{eqnarray}
converts the system of {\it differential} KS-DFT
equations into a system of {\it integro-differential} ones.
The solution of such a system of integro-differential equations is
more tedious procedure compared to a solution of KS-DFT equations with local
multiplicative potential, which is the same for {\it all}
KS orbitals. $E_{\rm x}^{\rm HF}$ in Eq.~(\ref{I1}) is the exact exchange energy
(the exchange energy of KS determinant) defined by the Hartree-Fock expression:
\begin{eqnarray}
E_{\rm x}^{\rm HF}[\{\phi_i\}]=
&-&
\frac{1}{2}
\sum_{i,j=1}^N \delta(\sigma_i, \sigma_j)
\nonumber\\
&\times&
\int\int
\frac{\phi_i^*({\bf r})\phi_j({\bf r})
\phi_j^*({\bf r}')\phi_i({\bf r}')}
{ |{\bf r}-{\bf r}'| }
d{\bf r}d{\bf r}'.
\label{I2}
\end{eqnarray}

To alleviate the non-locality of the HF exchange operator, the exact-expression approximate exchange (EEAX)
methods were developed. They all share the following common
properties:\cite{LHYB} (i) the exchange energy expression is the exact one defined by
Eq.~(\ref{I2}), and (ii) corresponding exchange potential is a local, multiplicative
operator. Among the first EEAX approximations, we should mention method introduced
by Krieger, Li, and Iafrate (KLI),\cite{KLI} the localized Hartree-Fock (LHF)
method and the common energy denominator approach (CEDA) introduced by
G\"orling\cite{Goerling.5718.01} and Gritsenko and Baerends\cite{Gritsenko.Baerends.CEDA} respectively. Later, self-consistent $\alpha$ (SC$\alpha$)\cite{SC1,SC2,KL01} and asymptotically-adjusted self-consistent
$\alpha$ (AASC$\alpha$)\cite{KL01a} local exchange functionals were
developed. Since the exchange energy in all of these methods is the exact one,
defined by Eq.~(\ref{I2}), the {\it self-interaction} error is eliminated, while multiplicative (or local in that sense)
exchange potential guarantees that these methods are fully in the framework of
DFT formalism. Another important aspect is that the EEAX methods
satisfy the Levy-Perdew virial relation.\cite{LevyPerdew85,ZhaoLevy93}

A solution to the problem of non-local operator in hybrid functionals
was proposed and implemented for the first time in Ref.\cite{LHYB},
when new local \lq\lq hybrid" exchange-correlation functionals
obtained by replacement
of the nonlocal Hartree-Fock (HF) exchange in conventional hybrid models by
one of the EEAX methods (the SC$\alpha$ and the AASC$\alpha$ approximations
are used in Ref.\cite{LHYB}).
Such functionals are fully within the framework of DFT formalism
because the functional derivative of EEAX functionals w.r.t. electronic density
is a local multiplicative potential. From the computational point of view,
the local \lq\lq hybrid" functionals can be viewed as more efficient
alternative to the conventional  hybrid methods.

The same scheme was later used to constract local \lq\lq hybrid" functionals
based on replacement the HF exchange in non-local hybrids
by the LHF approximation with local exchange potential.\cite{Goerling.115.04,Teale.Tozer.04,Arbuznikov.Kaupp.04}


A comparison  between non-local and local hybrids based on
the SC$\alpha$ and  AASC$\alpha$ methods which has involved
total and HOMO energies, vibrational frequencies and bond lengths
was made for diatomic molecules in Ref.\cite{LHYB}.
The comparison of other properties, in particular electric moments can
afford a more detailed measure
of the closeness of local and nonlocal hybrid functional performance.

In the present work we apply the local \lq\lq hybrid" functionals
introduced in \cite{LHYB}
for finite-difference (FD) calculations of total and HOMO energies, and
dipole and quadrupole electric moments for selected diatomic molecules.
To compare the
local \lq\lq hybrid" functional results with corresponding nonlocal
hybrid calculations, for which the basis set (BS) technique is used,
an estimate of basis set error (BSE) for each magnitude compared is needed.
For this purpose BSEs are calculated as the difference between the FD and BS results for the HF
and conventional DFT exchange-correlation functionals.

In Sec. II the local \lq\lq hybrid" functionals used in calculations
are introduced
and definitions of  the calculated quantities are established.
Numerical results are discussed in Sec. III. In Sec. III A the BSEs for total
energy and electric moments  of
standard basis sets available in GAUSSIAN  package\cite{Gaussian,Gaussian03}
are presented.
Local \lq\lq hybrid" functional self-consistent results
for total energy, multipole moments
and one-electron energies are compared to the values obtained from
calculations with  nonlocal
ones in Sec. III B.
Finally,  our conclusions are presented in Sec. IV.

\section{Method}


Two nonlocal hybrid functionals,  PBE0\cite{Adamo.Barone.1999} based
on the Perdew-Burke-Ernzerhof (PBE) DFT exchange-correlation functional,\cite{Perdew.Burke.Ernzerhof.1996}
and one parameter Becke-Lee-Yang and Parr hybrid, B1LYP,\cite{AdamoBarone97,Becke1040,Becke88,LYP} are used
to construct local \lq\lq hybrids".
The  nonlocal HF exchange in these non-local hybrids was replaced by
one of the EEAX term.
First, by SC$\alpha$-PBE and SC$\alpha$-B88 functionals
respectively (see Ref.\cite{KL01} for details),
and second, by AASC$\alpha$ exchange
(AA-m2 model from Ref.\cite{KL01a}).
For example, local SC-B1LYP \lq\lq hybrid" obtained from non-local
B1LYP functional by replacement of the non-local exchange term by
the local SC$\alpha$-B88 functional has the following form
\begin{eqnarray}
E_{\rm xc}^{\rm SC-B1LYP}=&a_0& E_{\rm x}^{\rm SC\alpha-B88}
+(1-a_0) (E_{\rm x}^{\rm LSD}+\Delta E_{\rm x}^{\rm B88})
\nonumber\\
&+&
E_{\rm c}^{\rm LSD}+\Delta E_{\rm c}^{\rm LYP},
\label{M1}
\end{eqnarray}
with $a_0=0.25$.
Expression of $E_{\rm x}^{\rm SC\alpha-B88}$ exchange functional is exact one
defined by Eq.~(\ref{I2}),
while exchange-correlation potential corresponding to functional of Eq.~(\ref{M1}) is a local
multiplicative operator
\begin{eqnarray}
v_{\rm xc}^{\rm SC-B1LYP}=&a_0& v_{\rm x}^{\rm SC\alpha-B88}+
(1-a_0)(v_{\rm x}^{\rm LSD}+\Delta v_{\rm x}^{\rm B88})
\nonumber\\
&+&
v_{\rm c}^{\rm LSD}
+\Delta v_{\rm c}^{\rm LYP},
\label{M2}
\end{eqnarray}
where potential $v_{\rm x}^{\rm SC\alpha-B88}$ is defined in Ref.\cite{KL01}.
On the whole, four local \lq\lq hybrids",
SC-PBE0, AAm2-PBE0 and SC-B1LYP, AAm2-B1LYP
were employed for numerical calculations of total and HOMO energies and
electric moments of selected diatomic molecules.

The multipole moments were calculated according to the following definition
(see Refs.\cite{McLeanYoshimine.1967,Kobus.PRA.2000}).
Equation for the independent component for each electric moment for a
system with axial symmetry
\begin{equation}
M^{(k)}=\sum_A Z_A R_{Az}-\int r^kP_k(z/r)  \rho({\bf r}) d^3{\bf r},
\label{M3}
\end{equation}
where $P_k$ are Legendre polynomials of degree $k$,
was used in finite-difference code
to calculate the first two moments
\begin{equation}
M^{(1)}\equiv\mu=\mu_z,
\label{M4}
\end{equation}
and
\begin{equation}
M^{(2)}\equiv\Theta=\Theta_{zz}.
\label{M5}
\end{equation}

The multipole moments were determined from the basis set calculations,
\begin{equation}
\mu_{\alpha}= \sum_A Z_A R_{A\alpha}-\int r_{\alpha}  \rho({\bf r}) d^3{\bf r},
\label{M6}
\end{equation}
and
\begin{equation}
Q_{\alpha\beta}= \sum_A Z_A R_{A\alpha}R_{A\beta}-\int r_{\alpha}r_{\beta}
\rho({\bf r}) d^3{\bf r},~~ (\alpha,\beta=x,y,z),
\label{M7}
\end{equation}
They can be related to components of Eqs.~(\ref{M3})-(\ref{M4}) evaluated by the
FD code (see Ref.\cite{Kobus.PRA.2000}):
\begin{equation}
\Theta_{zz}=Q_{zz}-{1\over2}(Q_{xx}+Q_{yy}).
\label{M8}
\end{equation}
Relations for higher moments can be found in Ref.\cite{Kobus.PRA.2000}.
In Eqs.~(\ref{M3}), (\ref{M6})-(\ref{M7}), $Z_A$ is the charge of nucleus $A$
and $R_{\alpha}$'s are its Cartesian coordinates and $r_{\alpha}$'s
are ($x,y,z$) coordinates.

The moments $M^{(1)}$ and $M^{(2)}$ in finite-difference calculations and
$\mu_z$, $\Theta_{zz}$ in basis set calculations were determined with
respect to the molecular center of mass (GAUSSIAN keyword Symmetry=CenterOfMass).


\section{Results and discussion}

\subsection{Basis set error for the HF and DFT methods}

Local \lq\lq hybrid" exchange-correlation functionals discussed in the
present work have been implemented in the spin-restricted FD full-numerical
program by Kobus, Laaksonen and Sundholm,\cite{KLSS96}
while the non-local hybrid functional calculations were performed using basis set based GAUSSIAN
package.\cite{Gaussian,Gaussian03}  The accuracy achieved in FD
calculations is about few $\mu$Hartree. The error of basis set truncation
is usually higher.

\begin{figure}[th]
  \centerline{\psfig{figure=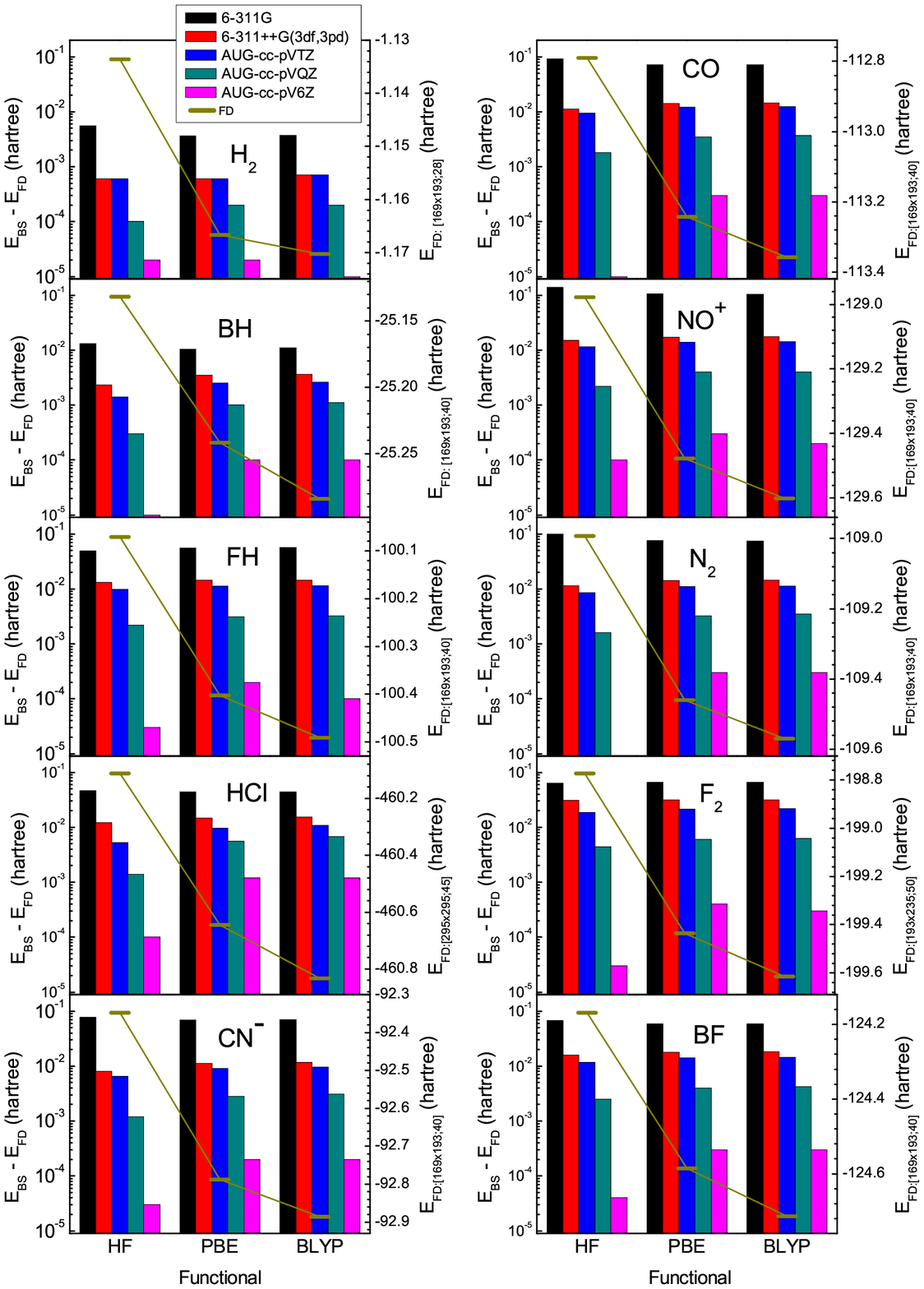,width=5.4in}}
  \caption{BSE for various basis sets (left axis, vertical bars) and total FD energies (right axis, horizontal bars connected to guide the eye) vs. functional used in HF and PBE, BLYP DFT calculations for selected diatomic molecules containing first-row atoms. BSE stands for the basis set error in total energy estimated by difference $E_{\rm Total}[{\rm BS}]-E_{\rm Total}[{\rm FD}]$, FD stands for the finite-difference calculations. \label{Fig1}}
\end{figure}

\begin{table}[th]
\caption{
Orbital energies ($-\varepsilon_i$) of the CO molecule
from the finite-difference  HF and  PBE, BLYP DFT calculations
and BSE$^a$ for different basis set calculations
(in Hartrees).
\label{Tab3} }
{\begin{tabular}{llrrr} \toprule
system & method &
$\varepsilon^{\rm HF}_i$/BSE &
$\varepsilon^{\rm PBE}_i$/BSE &
$\varepsilon^{\rm BLYP}_i$/BSE   \\
\colrule
1$\sigma$ & FD:  [169x193;40]        & 20.6645 & 18.8576  & 18.8807 \\
          & 6-311G                   & -0.0192 & -0.0045  & -0.0034 \\
          & 6-311++G(3df,3pd)        & -0.0032 &  0.0031  & +0.0028 \\
          & AUG-cc-pV6Z              &  0.0000 &  0.0002  & +0.0002\\
\colrule
2$\sigma$ & FD:  [169x193;40]        & 11.3600 &  9.9997  & 10.0249 \\
          & 6-311G                   & -0.0320 & -0.0073  & -0.0062 \\
          & 6-311++G(3df,3pd)        & -0.0009 &  0.0043  & +0.0038 \\
          & AUG-cc-pV6Z              &  0.0000 &  0.0001  & +0.0002 \\
\colrule
3$\sigma$ & FD:  [169x193;40]        & 1.5216  &  1.0769  & 1.0700 \\
          & 6-311G                   &-0.0427  & -0.0416  &-0.0395 \\
          & 6-311++G(3df,3pd)        &-0.0013  & -0.0010  &-0.0010 \\
          & AUG-cc-pV6Z     &$-1\cdot 10^{-5}$ &$5\cdot 10^{-6}$& $ +1\cdot 10^{-5}$\\
\colrule
4$\sigma$ & FD:  [169x193;40]        & 0.8045  &  0.5192  & 0.5144 \\
          & 6-311G                   &+0.0044  &  0.0095  &+0.0111 \\
          & 6-311++G(3df,3pd)        &-0.0007  & -0.0005  &-0.0005 \\
          & AUG-cc-pV6Z     &$-1\cdot 10^{-5}$ &$1\cdot 10^{-5}$&$+1\cdot 10^{-5}$\\
\colrule
1$\pi$ & FD:  [169x193;40]           & 0.6405  &  0.4360  & 0.4307 \\
          & 6-311G                   &-0.0059  & -0.0066  &-0.0056 \\
          & 6-311++G(3df,3pd)        &-0.0005  & -0.0002  &-0.0003 \\
          & AUG-cc-pV6Z              & 0.0000  &$5\cdot 10^{-6}$& -0.0001\\
\colrule
5$\sigma$ & FD:  [169x193;40]        & 0.5549  &  0.3322  & 0.3311 \\
          & 6-311G                   &-0.0023  &  0.0059  &+0.0071 \\
          & 6-311++G(3df,3pd)        &-0.0004  & -0.0001  &-0.0002 \\
          & AUG-cc-pV6Z              & 0.0000  &$8\cdot 10^{-6}$& $+1\cdot 10^{-5}$\\
\botrule
\end{tabular}}
\tablenotetext[1] {BSE stands for basis set error for orbital energy estimated by difference $\varepsilon_{i}[\rm BS]-\varepsilon_{i}[\rm FD]$.}
\end{table}

\begin{table}[th]
\caption{
Dipole moments  for the CO, BF and NO$^+$ diatomics
from the finite-differences and basis set HF and DFT calculations
and BSE$^a$ for various basis sets.
\label{Tab4}}
{\begin{tabular}{llrrrrrr} \toprule
system & numerical method &
\multicolumn{2}{c}{HF} &
\multicolumn{2}{c}{PBE} &
\multicolumn{2}{c}{BLYP}\\\cline{3-4}\cline{5-6}\cline{7-8}
&&
$\mu$ & BSE &
$\mu$ & BSE &
$\mu$ & BSE  \\
\colrule
CO       & BS: 6-311G           & -0.4974 & -0.2335  &0.1381  & -0.0864 & 0.1049 & -0.0806 \\
         & BS: 6-311++G(3df,3pd)& -0.2708 & -0.0069  &0.2213  & -0.0032 & 0.1811 & -0.0044 \\
         & BS: AUG-cc-pVTZ      & -0.2659 & -0.0020  &0.2238  & -0.0007 & 0.1845 & -0.0010 \\
         & BS: AUG-cc-pVQZ      & -0.2642 & -0.0003  &0.2243  & -0.0002 & 0.1853 & -0.0002 \\
         & BS: AUG-cc-pV6Z      & -0.2639 &  0.0000  &0.2246  & +0.0001 & 0.1857 & +0.0002 \\
         & FD: [169x193;40]     & -0.2639 &   --     &0.2245  &   --    & 0.1855 &   --    \\
\colrule
BF       & BS: 6-311G           & 0.4338  & -0.4410  &0.8692  & -0.2028 & 0.8379 & -0.1952 \\
         & BS: 6-311++G(3df,3pd)& 0.8740  & -0.0008  &1.0674  & -0.0046 & 1.0305 & -0.0026 \\
         & BS: AUG-cc-pVTZ      & 0.8764  & +0.0016  &1.0758  & +0.0038 & 1.0349 & +0.0018 \\
         & BS: AUG-cc-pVQZ      & 0.8744  & -0.0003  &1.0727  & +0.0007 & 1.0335 & +0.0004 \\
         & BS: AUG-cc-pV6Z      & 0.8748  &  0.0000  &1.0722  & +0.0002 & 1.0333 & +0.0003 \\
         & FD: [169x193;40]     & 0.8748  &   --     &1.0720  &   --    & 1.0331 & --      \\
\colrule
NO$^+$   & BS: 6-311G           & -0.8157 & -0.1750  &-0.4066 & -0.0927 & -0.4214 &-0.0925 \\
         & BS: 6-311++G(3df,3pd)& -0.6439 & -0.0032  &-0.3152 & -0.0013 & -0.3306 &-0.0017 \\
         & BS: AUG-cc-pVTZ      & -0.6388 & +0.0019  &-0.3121 &+0.0018  & -0.3269 & 0.0020 \\
         & BS: AUG-cc-pVQZ      & -0.6405 & +0.0002  &-0.3136 &+0.0003  & -0.3286 & 0.0003 \\
         & BS: AUG-cc-pV6Z      & -0.6408 & -0.0001  &-0.3139 & 0.0000  & -0.3288 &+0.0001 \\
         & FD: [169x193;40]     & -0.6407 &   --     &-0.3139 &   --    & -0.3289 &  --    \\
\botrule
\end{tabular}}
\tablenotetext[1] {BSE stands for basis set error for the moments estimated by differences $\mu[\rm BS]-\mu[\rm FD]$.}
\end{table}

On Figure~\ref{Fig1}
the total energies of finite-difference calculations
for the HF method and for the PBE and BLYP functional DFT calculations
are presented. These values are compared to the basis set results for
different basis sets. Corresponding BSEs defined as difference
between FD and BS total energies are plotted. As one would expect,
these values, due to the variational principle, are always positive
(a numerical error of the FD code is smaller than the basis set
truncation error).
Values corresponding to the
6-311G basis set are quite similar for the HF and DFT methods and they
are relatively small for the H$_2$ and BH molecules
(an order of 0.005-0.01 Hartrees).
These errors increase for other systems up to
0.05-0.1 Hartrees. Augmentation of 6-311G basis with polarization and diffuse functions
decreases an error by the factor of 2 to 10 for the molecules
presented on Figure~\ref{Fig1} 
(factor of 2 for the F$_2$ molecule
and factor of ~10 for the CN$^-$, CO, NO$^+$ and N$_2$ systems).

Finally, the BSEs of the largest AUG-cc-pV6Z 
basis sets are in fourth decimal place in most cases (with few
exceptions, where the value is smaller). The AUG-cc-pV6Z basis set error in
DFT calculations for the HCl molecule is larger (0.0012 Hartrees).

As one can see from the Figure~\ref{Fig1}, 
the errors in HF method
is smaller than the errors in DFT calculations
when the AUG-cc-pV(T,Q,6)Z basis sets are employed.

The errors introduced by incomplete basis set in one-electron
orbital energies are presented in Table~\ref{Tab3} for CO molecule as an example.
The BSEs for the 6-311G basis are in third or in second
decimal place. The addition of polarization and diffuse functions decrease
these errors significantly.
The AUG-cc-pV6Z basis set provides practically exact values to within four
decimal places.

Convergence of the dipole moment values with the basis set
(and corresponding BSEs) for three molecules, CO, BF and NO$^+$,
are shown in Table~\ref{Tab4}.
The basis set error decreases very fast with increasing the basis set size.
For the largest available basis set the moments are reproduced quantitatively
within four decimal places.
The same trends are observed for the quadrupole moment values, shown in Table~\ref{Tab5}.
The convergence is fast for all the molecules presented in Table~\ref{Tab5}. The error for
the largest basis set employed (AUG-cc-pV6Z) is in fourth decimal.

In Table~\ref{Tab6} we present the dipole and quadrupole moments from FD and
largest BS calculations corresponding to the HF
and the DFT methods. The BSEs are calculated as
$\mu[\rm BS]-\mu[\rm FD]$, and $\Theta[\rm BS]-\Theta[\rm FD]$.
The AUG-cc-pV6Z basis set errors for dipole moment $\mu$ in most cases
are in fourth decimal except for
the DFT calculations of the HCl molecule and the CN$^-$
anion, where the errors are 0.0012 D and 0.0049 D respectively.

The quadrupole moment basis set errors are also reproduced within four decimal places with
a few exceptions. DFT calculations for the CN$^-$ ion shows the largest
error (0.0138-0.0150 D$\AA$).

Here we should emphasize that the numerical errors of the finite-difference code
for the multipole moments are in the order of $1\cdot 10^{-6}$ D for the dipole moment $\mu$
and in the order of $1\cdot 10^{-7}$ D$\AA$ for the quadrupole moment $\Theta$. These values
are much smaller than differences presented in Table \ref{Tab6} which, consequently,
can be considered as basis set errors corresponding to the multipole
moments $\mu$ and $\Theta$.

\begin{table}
\caption{
Quadrupole  moments  for the CO, BF and NO$^+$ diatomics
from the finite-differences and basis set HF and DFT calculations
and BSE$^a$ for various basis sets.
\label{Tab5}}
{\begin{tabular}{llrrrrrr} \toprule
system & numerical method &
\multicolumn{2}{c}{HF} &
\multicolumn{2}{c}{PBE} &
\multicolumn{2}{c}{BLYP}\\\cline{3-4}\cline{5-6}\cline{7-8}
&&
$\Theta$ & BSE &
$\Theta$ & BSE &
$\Theta$ & BSE  \\
\colrule
CO       & BS: 6-311G           & -2.8917 & -0.8332 & -2.6523 & -0.6389 & -2.6637 & -0.6165 \\
         & BS: 6-311++G(3df,3pd)& -2.1252 & -0.0667 & -2.0769 & -0.0635 & -2.1071 & -0.0599 \\
         & BS: AUG-cc-pVTZ      & -2.0799 & -0.0214 & -2.0370 & -0.0236 & -2.0698 & -0.0226 \\
         & BS: AUG-cc-pVQZ      & -2.0590 & -0.0005 & -2.0181 & -0.0047 & -2.0517 & -0.0045 \\
         & BS: AUG-cc-pV6Z      & -2.0588 & -0.0003 & -2.0138 & -0.0004 & -2.0478 & -0.0006 \\
         & FD: [169x193;40]     & -2.0585 &   -     & -2.0134 & -       & -2.0472 &   -     \\
\colrule
BF       & BS: 6-311G           & -4.8487 & -0.6031 & -3.8964 & -0.5340 & -3.9176 & -0.5072 \\
         & BS: 6-311++G(3df,3pd)& -4.2418 &  0.0038 & -3.3513 &  0.0111 & -3.4025 & 0.0079  \\
         & BS: AUG-cc-pVTZ      & -4.2422 &  0.0034 & -3.3689 & -0.0065 & -3.4178 & -0.0074 \\
         & BS: AUG-cc-pVQZ      & -4.2417 &  0.0039 & -3.3662 & -0.0038 & -3.4137 & -0.0033 \\
         & BS: AUG-cc-pV6Z      & -4.2449 &  0.0000 & -3.3627 & -0.0003 & -3.4107 & -0.0004 \\
         & FD: [169x193;40]     & -4.2456 & -       & -3.3624 & -       & -3.4104 & -       \\
\colrule
NO$^+$   & BS: 6-311G           & -0.2997 & -0.9906 & -0.2349 & -0.8152 & -0.2499 & -0.7993 \\
         & BS: 6-311++G(3df,3pd)&  0.6984 &  0.0075 &  0.5793 & -0.0010 &  0.5509 & 0.0015  \\
         & BS: AUG-cc-pVTZ      &  0.6926 &  0.0017 &  0.5769 & -0.0034 &  0.5471 & -0.0023 \\
         & BS: AUG-cc-pVQZ      &  0.6946 &  0.0037 &  0.5819 &  0.0016 &  0.5513 & 0.0019  \\
         & BS: AUG-cc-pV6Z      &  0.6908 & -0.0001 &  0.5800 & -0.0003 &  0.5492 & -0.0002 \\
         & FD: [169x193;40]     &  0.6909 &     -   &  0.5803 &   -     &  0.5494 &  -      \\
\botrule
\end{tabular}}
\tablenotetext[1] {BSE stands for basis set error for the moments estimated by differences $\Theta[\rm BS]-\Theta[\rm FD]$.}
\end{table}

\begin{table}
\caption{
Multipole moments  for selected diatomic molecules
from the basis set HF and PBE, BLYP DFT calculations
and comparison to the finite-difference results.
\label{Tab6}}
{\begin{tabular}{llrrrrrr} \toprule
system & numerical method &
\multicolumn{2}{c}{HF} &
\multicolumn{2}{c}{PBE} &
\multicolumn{2}{c}{BLYP}\\\cline{3-4}\cline{5-6}\cline{7-8}
&&
$\mu$ & $\Theta$ &
$\mu$ & $\Theta$ &
$\mu$ & $\Theta$  \\
\colrule
H$_2$    & BS: AUG-cc-pV6Z      & 0.0     & 0.6640  & 0.0   & 0.5786& 0.0 & 0.5658 \\
         & FD: [169x193;28]     & 0.0     & 0.6639  & 0.0   & 0.5785& 0.0 & 0.5658 \\
         & BSE$^a$              & 0.0     &+0.0001  & 0.0   &+0.0001& 0.0 & 0.0000 \\
\colrule
BH       & BS: AUG-cc-pV6Z      & -1.7301 & -3.6021 &-1.5261&-3.2986& -1.5120 & -3.3423 \\
         & FD: [169x193;40]     & -1.7301 & -3.6017 &-1.5259&-3.2986& -1.5118 & -3.3413 \\
         & BSE                  & +0.0000 & -0.0004 &-0.0002& 0.0000& -0.0002 & -0.0010 \\
\colrule
FH       & BS: AUG-cc-pV6Z      & -1.9219 & 2.3300  &-1.7457& 2.2564& -1.7511 &  2.2503 \\
         & FD: [193x235;50]     & -1.9218 & 2.3299  &-1.7458& 2.2566& -1.7510 &  2.2506 \\
         & BSE                  & -0.0001 &+0.0001  &+0.0001&-0.0002& -0.0001 & -0.0003 \\
\colrule
HCl      & BS: AUG-cc-pV6Z      &-1.1797  & 3.7679  &-1.0695& 3.5819&-1.0474 & 3.5473 \\
         & FD: [295x295;45]     &-1.1790  & 3.7664  &-1.0683& 3.5803&-1.0463 & 3.5453 \\
         & BSE                  &-0.0007  &+0.0015  &-0.0012&+0.0016&-0.0011 &+0.0020 \\
\colrule
CN$^-$   & BS: AUG-cc-pV6Z      & 0.3928  &-4.4613  & 0.7138&-5.0930& 0.6767 &-5.1122\\
         & FD: [169x193;40]     & 0.3929  &-4.4608  & 0.7177&-5.0792& 0.6816 &-5.0972\\
         & BSE                  &-0.0001  &-0.0005  &-0.0039&-0.0138&-0.0049 &-0.0150\\
\colrule
NO$^+$   & BS: AUG-cc-pV6Z      &-0.6408  & 0.6908  &-0.3139& 0.5800&-0.3288 & 0.5492\\
         & FD: [169x193;40]     &-0.6407  & 0.6909  &-0.3139& 0.5803&-0.3289 & 0.5494\\
         & BSE                  &-0.0001  &-0.0001  & 0.0000&-0.0003&+0.0001 &-0.0002\\
\colrule
N$_2$    & BS: AUG-cc-pV6Z      & 0.0     &-1.2507  & 0.0   &-1.5359& 0.0    &-1.5660\\
         & FD: [169x193;40]     & 0.0     &-1.2509  & 0.0   &-1.5364& 0.0    &-1.5663\\
         & BSE                  & 0.0     &+0.0002  & 0.0   &+0.0005& 0.0    &+0.0003\\
\colrule
F$_2$    & BS: AUG-cc-pV6Z      & 0.0     & 0.6758  & 0.0   & 0.8994& 0.0    & 0.8559\\
         & FD: [193x235;50]     & 0.0     & 0.6753  & 0.0   & 0.8994& 0.0    & 0.8558\\
         & BSE                  & 0.0     &+0.0005  & 0.0   & 0.0000& 0.0    &+0.0001\\
\botrule
\end{tabular}}
\tablenotetext[1] {BSE stands for basis set error for the moments estimated by differences $\mu[\rm BS]-\mu[\rm FD]$ and $\Theta[\rm BS]-\Theta[\rm FD]$.}
\end{table}

\subsection{Local \lq\lq hybrid" functional calculations: comparison to non-local hybrids}

Total energies, dipole and quadrupole moments for 10 diatomic molecules
obtained from
different exchange-correlation local \lq\lq hybrid" functional methods
are presented in Tables~\ref{Tab7}, \ref{Tab8}, and \ref{Tab9}. 
The differences between values obtained from
nonlocal hybrid functional calculations and values
corresponding to the local \lq\lq hybrid" functional are given in parenthesis.
For example, differences presented in parenthesis of the second and third
column of Table \ref{Tab7} are $\Delta=E_{\rm PBE0}[\rm BS]-E_{\rm SC-PBE0}[\rm FD]$
and $\Delta=E_{\rm PBE0}[\rm BS]-E_{\rm AAm2-PBE0}[\rm FD]$ correspondingly.

The BSEs  of basis sets employed for the nonlocal hybrid
functional calculations are given for comparison.
These values are estimated by difference between corresponding magnitudes
of the BS and the FD DFT calculations
and they are taken from Figure~\ref{Fig1} and Tables~\ref{Tab4}, \ref{Tab5}, and \ref{Tab6}. 

Comparison of $\Delta$'s with corresponding values of BSE shows that
total energies obtained from the  AAm2-PBE0 local \lq\lq hybrid" functional
calculations are coincide with energies coming from the nonlocal
PBE0 hybrid functional within the magnitude of BSE practically for all cases with
one exception for the F$_2$ molecule. For the SC-PBE0 local functional
the values of $\Delta$ are larger than corresponding
values of BSE's but they do not exceed 0.0022 Hartree.
The average absolute values of $\Delta$'s 
are 0.0012 and 0.0004 Hartree for
SC-PBE0 and AAm2-PBE0 local functionals correspondingly
and the average BSE for the PBE DFT functional has the value of 0.0003 Hartee.

The situation is quite
similar for the SC-B1LYP and AAm2-B1LYP local functionals:
the average absolute values of $\Delta$'s for the AAm2-B1LYP functional
are very close to the average
BSE (0.0005 and  0.0003 Hartree correspondingly). The maximum value of $\Delta$
for the SC-B1LYP functional is 0.0021 Hartree for the F$_2$ diatomic.

A different trend is observed for the electric multipole moment calculations.
Values of $\Delta$ which correpond to difference between non-local hybrids and local
functional values are approximately one order of magnitude larger
than corresponding values of BSE.
For example, the average BSE value of BLYP
functional for dipole moment is 0.0010 D, the average absolute
values of $\Delta$ for the SC-B1LYP and AAm2-B1LYP functionals
(i.e. average absolute differences between dipole moment values
corresponding to the non-local B1LYP hybrid and to the  local \lq\lq hybrid"
functionals) are much larger (0.0753 D for the SC-B1LYP and 0.0366 D for the
AAm2-B1LYP). Dipole moment mean absolute error (MAE) of local \lq\lq hybrid" functionals based on the
asymptotically-adjusted potential (AAm2-PBE0 and AAm2-B1LYP) is about
twice smaller as compared to the functionals based on the self-consistent $\alpha$
method (SC-PBE0 and SC-B1LYP). This could be explained by the fact that the
AAm2 model for the exchange potential much better mimics the non-local behavior
of the non-local Hartree-Fock exchange, that appears in conventional hybrids
(see details in \cite{LHYB,KL01,KL01a}).

In spite of fact that values of $\Delta$ for multipole moments
are larger than corresponding BSE values,
the relative discrepancies between local and non-local hybrid
functional values are reasonably small
(order of few percents for both dipole and quadrupole moments) ,
with exceptions for CN$^-$, CO and NO$^+$ dipole moment calculations
when SC-PBE0 and SC-B1LYP
local functionals are used (when the relative error is between 10-60 \%). 
Comparing MAE from Tables~\ref{Tab8} and \ref{Tab9} with BSE from Tables~\ref{Tab4} and \ref{Tab5} respectively, we conclude that the finite-difference method combined with local ``hybrid" functional produces dipole and quadrupole moments with better accuracy, than the corresponding non-local hybrid functional does in combination with the 6-311G basis set and worse than that with 6-311++G**.

One-electron energies corresponding to the six molecular orbitals of the
CO molecule obtained from the local and nonlocal hybrids are presented
in Table \ref{Tab10}. The values for the $1\sigma$ and $2\sigma$ inner orbitals
coincide for local and nonlocal methods within two significant figures
(corresponding differences are about 2-4 \% of orbital energy value).
For the $1\pi$ and $5\sigma$ outer orbitals the local functionals
AAm2-PBE0 and AAm2-B1LYP yield values quite close to those of
the nonlocal hybrid methods
(corresponding differences are about 2\% for the $1\pi$ and smaller than
1\% for the HOMO energy value). Whereas the SC$\alpha$ based local
\lq\lq hybrid" functional one-electron energies differ in second
significant figure (corresponding differences are about 16\% of
orbital energy magnitude).

The fact that the HOMO energies obtained from the
AAm2-based local \lq\lq hybrid" functional calculations are quite close
to those of the original nonlocal ones is explained by correct asymptotic form ($-1/r$)
of the  AAm2 exchange potential, whereas SC$\alpha$ exchange potential
has an asymptotic behavior similar to the corresponding DFT functional
(PBE and B88 exchange potentials in our case) (see Ref.\cite{LHYB} for details).


\begin{table}[th]
\caption{
Total energies from the finite-difference  SC/AAm2-PBE0 and SC/AAm2-B1LYP local \lq\lq hybrid" functional calculations
and comparison to the basis set PBE0 and B1LYP nonlocal  hybrid functional self-consistent results
(the differences between non-local hybrid and corresponding local \lq\lq hybrid" functional results,
$\Delta=E_{\rm non-local-hyb}^a-E_{\rm local-\lq\lq hyb"}$, are presented in parenthesis. All anergies are in Hartrees).
\label{Tab7}}
{\begin{tabular}{lrrrrrr} \toprule
system &
$E_{\rm SC-PBE0}(\Delta)$ &
$E_{\rm AAm2-PBE0}(\Delta)$ &
BSE$^b$ &
$E_{\rm SC-B1LYP}(\Delta)$ &
$E_{\rm AAm2-B1LYP}(\Delta)$ &
BSE$^c$ \\
\colrule
H$_2$    & -1.1689(-0.0002)& -1.1691(0.0000) &   -1.1691 ~0.0000 & -1.1703(-0.0001)& -1.1704( 0.0000)& -1.1704   ~0.0000 \\
BH       & -25.2491(-0.0004)& -25.2494(-0.0001)&  -25.2495 ~0.0001 & -25.2838(-0.0003)& -25.2840(-0.0001)& -25.2841  ~0.0001 \\
FH       & -100.4005(-0.0009)& -100.4013(-0.0001)& -100.4014 ~0.0002 & -100.4728(-0.0010)& -100.4736(-0.0002)& -100.4738 ~0.0001 \\
HCl      & -460.6797(-0.0003)& -460.6790(-0.0010)& -460.6800 ~0.0012 & -460.8301(-0.0004)& -460.8294(-0.0011)& -460.8305 ~0.0012 \\
CN$^-$   & -92.7807(-0.0015)& -92.7819(-0.0003)&  -92.7822 ~0.0002 & -92.8608(-0.0015)&  -92.8619(-0.0004)& -92.8623  ~0.0002 \\
CO       & -113.2363(-0.0015)& -113.2374(-0.0004)& -113.2378 ~0.0003 & -113.3312(-0.0014)& -113.3322(-0.0004)& -113.3326 ~0.0003 \\
NO$^+$   & -129.4622(-0.0015)& -129.4632(-0.0005)& -129.4637 ~0.0003 & -129.5637(-0.0014)& -129.5646(-0.0005)& -129.5651 ~0.0002 \\
N$_2$    & -109.4514(-0.0013)& -109.4524(-0.0003)& -109.4527 ~0.0003 & -109.5408(-0.0013)& -109.5417(-0.0004)& -109.5421 ~0.0003 \\
F$_2$    & -199.4165(-0.0022)& -199.4180(-0.0007)& -199.4187 ~0.0004 & -199.5644(-0.0021)& -199.5657(-0.0008)& -199.5665 ~0.0003 \\
BF       & -124.5866(-0.0016)& -124.5877(-0.0005)& -124.5882 ~0.0003 & -124.6930(-0.0016)& -124.6940(-0.0006)& -124.6946 ~0.0003 \\
\colrule
MAE & (0.0012) & (0.0004) & 0.0003 & (0.0011) & (0.0005) & 0.0003\\
\botrule
\end{tabular}}
\tablenotetext[1] {Nonlocal hybrid functional calculations were performed using the largest basis sets presented in Figure~\ref{Fig1}.}
\tablenotetext[2] {Basis set error estimated by difference $E_{\rm PBE}[\rm BS]-E_{\rm PBE}[\rm FD]$ is taken from Figure~\ref{Fig1}.}
\tablenotetext[3] {Basis set error estimated by difference $E_{\rm BLYP}[\rm BS]-E_{\rm BLYP}[\rm FD]$ is taken from Figure~\ref{Fig1}.}
\end{table}

\begin{table}[th]
\caption{
Dipole moments from the finite-difference SC/AAm2-PBE0 and SC/AAm2-B1LYP
local \lq\lq hybrid" functional calculations and comparison to the  basis set PBE0 and B1LYP
nonlocal  hybrid functional$^a$ self-consistent results.
(Differences between non-local hybrid
and corresponding local \lq\lq hybrid" functional results,
$\Delta=\mu_{\rm non-local-hyb}^a-\mu_{\rm local-\lq\lq hyb"}$,
are presented in parenthesis).
\label{Tab8}}
{\begin{tabular}{lrrrrrr} \toprule
system &  SC-PBE0 &  AAm2-PBE0 &
BSE$^b$ &
SC-B1LYP & AAm2-B1LYP &
BSE$^b$ \\
\colrule
BH      &-1.5248(-0.0555)&-1.5379(-0.0424) & -1.5803~-0.0002 &-1.5119(-0.0548) &-1.5254(-0.0413) & -1.5667 ~ -0.0002 \\
FH       &-1.7457(-0.0560) &-1.8427(+0.0410) & -1.8017~+0.0001 &-1.7503(-0.0572) &-1.8485(+0.0410) & -1.8075 ~ -0.0001 \\
HCl      &-1.0686(-0.0430) &-1.1376(+0.0260) & -1.1116~-0.0012 &-1.0463(-0.0438) &-1.1170(+0.0269) & -1.0901 ~ -0.0011 \\
CN$^-$   & 0.7183(-0.0888) & 0.6159(+0.0136) &  0.6295~-0.0039 & 0.6836(-0.0923) & 0.5805(+0.0108) & 0.5913  ~ -0.0049 \\
CO       & 0.2250(-0.1225) & 0.1090(-0.0065) &  0.1025~+0.0001 & 0.1875(-0.1251) & 0.0715(-0.0091) & 0.0624  ~ +0.0002 \\
NO$^+$   &-0.3135(-0.0767) &-0.3673(-0.0229) & -0.3902~+0.0000 &-0.3277(-0.0782) &-0.3816(-0.0243) & -0.4059 ~ +0.0001 \\
BF       & 1.0719(-0.0713) & 0.8967(+0.1039) &  1.0006~+0.0002 & 1.0350(-0.0758) & 0.8565(+0.1027) & 0.9592  ~ +0.0003 \\
\colrule
MAE      &       (0.0734) &         (0.0366)    &   0.0008   &    (0.0753)  &    (0.0366) & 0.0010\\
\botrule
\end{tabular}}
\tablenotetext[1] {Nonlocal hybrid functional calculations of the electric moments were performed using the largest basis sets presented in Figure~\ref{Fig1}.}
\tablenotetext[2] {Basis set error is estimated by values taken from Tables \ref{Tab4}, \ref{Tab6}.}
\end{table}

\begin{table}[th]
\caption{
Quadrupole moments from the finite-difference
SC/AAm2-PBE0 and SC/AAm2-B1LYP
local \lq\lq hybrid" functional calculations
and comparison to the  basis set PBE0 and B1LYP
nonlocal  hybrid functional$^a$ self-consistent results.
\label{Tab9}}
{\begin{tabular}{lrrrrrr} \toprule
system &  SC-PBE0 &  AAm2-PBE0 &
BSE$^b$ &
SC-B1LYP & AAm2-B1LYP &
BSE$^b$ \\
\colrule
H$_2$    &   0.5787 (+0.0282)& 0.6069 ( 0.0000)&  0.6069~ +0.0001 & 0.5658 (+0.0291)& 0.5949 (+0.0000)& 0.5949  ~ +0.0000 \\
BH       &  -3.2961 (-0.0358)&-3.2956 (-0.0363)& -3.3319~  0.0000 &-3.3414 (-0.0315)&-3.3369 (-0.0360)& -3.3729 ~ -0.0010 \\
FH       &   2.2566 (+0.0217)& 2.3152 (-0.0369)&  2.2783~ -0.0002 & 2.2507 (+0.0230)& 2.3115 (-0.0378)&  2.2737 ~ -0.0003 \\
HCl      &   3.5797 (+0.0603)& 3.6362 (+0.0038)&  3.6400~ +0.0016 & 3.5453 (+0.0622)& 3.6031 (+0.0044)&  3.6075 ~ +0.0020 \\
CN$^-$   &  -5.0803 (+0.2254)&-5.0129 (+0.1580)& -4.8549~ -0.0138 &-5.1011 (+0.2362)&-5.0321 (+0.1672)& -4.8649 ~ -0.0150 \\
CO       &  -2.0136 (+0.0162)&-2.0219 (+0.0245)& -1.9974~ -0.0004 &-2.0481 (+0.0205)&-2.0538 (+0.0262)& -2.0276 ~ -0.0006 \\
NO$^+$   &   0.5801 (+0.0421)& 0.5742 (+0.0480)&  0.6222~ -0.0003 & 0.5489 (+0.0441)& 0.5439 (+0.0491)&  0.5930 ~ -0.0002 \\
N$_2$    &  -1.5368 (+0.0925)&-1.5369 (+0.0926)& -1.4443~ +0.0005 &-1.5675 (+0.0960)&-1.5668 (+0.0953)& -1.4715 ~ +0.0003 \\
F$_2$    &   0.8994 (-0.0572)& 0.9360 (-0.0938)&  0.8422~  0.0000 & 0.8558 (-0.0563)& 0.8964 (-0.0969)&  0.7995 ~ +0.0001 \\
BF       &  -3.3624 (-0.2031)&-3.3712 (-0.1943)& -3.5655~ -0.0003 &-3.4102 (-0.1886)&-3.4048 (-0.1940)& -3.5988 ~ -0.0004 \\
\colrule
MAE      &          (0.0783) &       (0.0688) &     0.0017    &      (0.0787) &      (0.0707)  & 0.0020\\
\botrule
\end{tabular}}
\tablenotetext[1] {Nonlocal hybrid functional calculations of the electric moments were performed using the largest basis sets presented in Figure~\ref{Fig1}.}
\tablenotetext[2] {Basis set error is estimated by values taken from Tables~\ref{Tab5}, \ref{Tab6}.}
\end{table}

\begin{table}[th]
\caption{
Orbital energies energies ($-\varepsilon_i$) of the CO molecule
from the finite-difference  SC/AAm2-PBE0 and SC/AAm2-B1LYP
local \lq\lq hybrid" functional calculations
and comparison to the basis set PBE0 and B1LYP
nonlocal  hybrid functional self-consistent results
(all energies are in Hartrees).
\label{Tab10}}
{\begin{tabular}{lrrrrrr} \toprule
system & SC-PBE0 &
AAm2-PBE0 &
PBE0 &
SC-B1LYP &
AAm2-B1LYP &
B1LYP
   \\
\colrule
1$\sigma$ & 18.8566 & 18.9016 & 19.3144 & 18.8765 & 18.9216 & 19.3350 \\
2$\sigma$ &  9.9989 & 10.1093 & 10.3442 & 10.0215 & 10.1318 & 10.3675 \\
3$\sigma$ &  1.0767 &  1.1536 & 1.1976  &  1.0692 &  1.1469 &  1.1907 \\
4$\sigma$ &  0.5191 &  0.5910 & 0.5996  &  0.5137 &  0.5862 &  0.5947 \\
1$\pi$    &  0.4358 &  0.5048 & 0.4961  &  0.4301 &  0.4997 &  0.4907 \\
5$\sigma$ &  0.3321 &  0.3948 & 0.3950  &  0.3304 &  0.3935 &  0.3939 \\
\botrule
\end{tabular}}
\end{table}

\section{Conclusions}

Local \lq\lq hybrid" functionals obtained by replacement of non-local Hartree-Fock exchange
in conventional hybrids by local EEAX demonstrate the performance close
to the original non-local hybrids.
Expression for the total energy is the same in both cases
(local and non-local hybrids), while the corresponding
KS potential for local hybrids is fully local and multiplicative operator that brings back the
local hybrids in the framework of density functional theory.

In the present work the local \lq\lq hybrid" functional results, total energies,
orbital energies, dipole and quadrupole moments were compared to conventional
hybrid functional values for selected
diatomic molecules. The total energies were found to coincide for both type of functionals within the
basis set error values. Multipole moments provide detailed information about
spatial distribution of electron density. Due to sensitivity of these moments to the
details of DFT calculation (functional
employed, basis set, etc.), the discrepancies between the local and non-local
hybrid functional results are much larger than corresponding basis set errors. The
typical percentage error is usually less than 10\% with a few exceptions.
Hence the local \lq\lq hybrid" functionals represent a computationally less expensive
alternative to the non-local hybrids
in framework of DFT and provide a local multiplicative exchange-correlation potential.
Properties related to total and orbital energies, and density distributions
obtained by local functionals were found in good agreement with those from
the original non-local hybrids.

\section{Acknowledgments}

V.V.K. acknowledges informative conversations with Samuel B. Trickey.
V.V.K. was supported by the US Department of Energy TMS program 
(Grant No. DE-SC0002139). 
I.A.M. and A.E.M. are grateful for NSF support via grant
CHE 0832622, and for use computer time allotted on DOE NERSC, and UCF
IST Stokes supercomputing facilities.

\end{document}